\documentstyle[12pt,epsf]{article}
\begin{document}

\title{\bf Is there a $\rho$ in the  O(4) $\lambda \phi^4_4$
theory?\footnote{Submitted to Physics Letters B.}}                      
\vskip 1. truein
\author{Khalil M. Bitar and Pavlos M. Vranas \\[3ex]
\em Supercomputer Computations Research Institute \\
\em The Florida State University \\
\em Tallahassee, FL 32306 }

\maketitle

\vskip 1.5 truein

\begin{abstract}

A Monte Carlo simulation of the $O(4)$ $\lambda \phi^4$ theory in the
broken phase is performed on a hypercubic lattice in search of an
$I=1$, $J=1$ resonance. The region of the
cutoff theory where the interaction is strong is investigated since
it is there that a resonance would be expected to have a better chance
to form. In that region the presence of an $I=1$, $J=1$ resonance with
mass below the cutoff is excluded.

\end{abstract}

\newpage

The question of whether or not an $I=1$, $J=1$ resonance exists in the
broken phase of the four component $\lambda \phi^4$ theory in four
dimensions is in fact an old one. In the sixties, a quite
phenomenological theory describing the low energy pion processes was
developed by Gell--Mann and Levy \cite{Gell--Mann&Georgi}.  The theory
is described by a chiral Lagrangian, and involves the three pion
fields $\pi^1, \pi^2, \pi^3$, the scalar field $\sigma$, the two
nucleons, their interactions, and the dimensionful pion decay constant
$f_\pi=90 \ MeV$. The $\sigma$, although it has never really been
seen, may exist in nature as a broad and quite heavy ($\approx 700 \
MeV$) isospin I=0, spin J=0 resonance.  In that setting, it was
natural to ask whether the presence of the $\rho$ resonance was a
consequence of the pion interactions or of some higher energy physics.
Since the theory of Gell--Mann and Levy becomes an O(4) $\lambda
\phi^4$ theory if the nucleon fields are neglected, this question
reduces to whether or not the O(4) $\lambda \phi^4$ theory can sustain
an $I=1$, $J=1$ resonance in the broken phase.  Earlier attempts
\cite{Basdevant&Lee} to answer this question used analytical
approximations such as Pade approximants and found that an $I=1$, $J=1$
resonance is present in the theory. It is not clear, however, whether
or not the presence of this resonance is an artifact of the
approximations used and therefore the question of the existence of an
$I=1$, $J=1$ resonance in the theory has yet to receive a conclusive
answer.

Today, in a very different setting, this same question is of interest
again.  The scalar sector of the Minimal Standard Model is also a four
component $\lambda \phi^4$ theory in the broken symmetry phase.  The
equivalent of the $\sigma$ resonance is the Higgs particle, the three
pions are the Goldstone bosons, the pion decay constant is the weak
scale $f_G=246 \ GeV$, and the scattering of longitudinally polarized
vector bosons behaves exactly the same way as a scaled up version of
$\pi$--$\pi$ scattering (equivalence theorem \cite{equivalence}).  It
is possible that the Higgs, like the $\sigma$, is quite heavy and
broad and may avoid detection at SSC. On the other hand, if the four
component $\lambda \phi^4$ theory does indeed contain an $I=1$, $J=1$
resonance, then since the $\rho$ resonance, as it appears in nature,
is quite strong, its equivalent in the scalar sector of the Minimal
Standard Model may have a good chance to be detected at SSC.
Therefore, it becomes very important to know if another ``signature''
of the scalar sector, besides the Higgs resonance, is awaiting
discovery at SSC.

A direct Monte Carlo simulation was decided to be the best way to shed
new light on this old question (a leading order large N calculation
was not able to produce an $I=1$, $J=1$ resonance).  The O(4) $\lambda
\phi^4$ theory was simulated on the lattice in the $\lambda
\rightarrow \infty$ limit. In that limit, the theory has the strongest
interactions and a resonance probably has a better chance to form.
The model has the lattice action
\begin{equation}
S=\sum_{x \in \Lambda}\left\{ 
-k \sum_{\mu=1}^{4} \left( \vec\Phi_x\vec\Phi_{x+\hat\mu}+
\vec\Phi_x\vec\Phi_{x-\hat\mu} \right) +
\lambda (\vec\Phi_x\vec\Phi_{x}-1)^2+ 
\vec\Phi_x\vec\Phi_{x}
\right\} \label{action}
\end{equation}
and was simulated on a hypercubic lattice $\Lambda$ of spatial
extension $L$ and time extension $L_t$ using an incomplete heat bath
algorithm \cite{Fredenhagen&Marcu} on the CM-2 machine at SCRI.

Since, on a finite lattice, the direction of the symmetry breaking
changes from configuration to configuration, the same approach as in
\cite{Has} was used to disentangle the massive scalar field $\sigma$
from the Goldstone modes.  For each configuration, the global O(4)
coordinate system was rotated so that its first axis would be parallel
to the direction of the symmetry breaking. Because there are many ways
to perform this rotation, a ``simple'' one was consistently used
throughout the simulation. The field, expressed in the new coordinate
system, is $\vec\Phi_x=(\sigma_x,\pi_x^1,\pi_x^2,\pi_x^3)$, with
$\sigma$ being the massive scalar field and $\pi^1,\pi^2,\pi^3$ the
three massless pion fields.

To measure the mass $m_\sigma$ of the $\sigma$ field, the time slice
connected correlation function
$C_{0,0}(t)=<O_\sigma(0),O_\sigma(t)>_c\ $ 
of the zero 3--momentum operator
\begin{equation}
O_\sigma(t)={1\over{L^3}} \sum_{x \in \Lambda_t}\sigma_x \ ,
\label{sigma_op}
\end{equation}
where $\Lambda_t$ is the three dimensional time slice at Euclidean
time $t$,
was fitted to 
\begin{equation}
C(t)=a\cosh\left[M\left(t-{L_t\over 2}\right)\right]+b
\label{cosh}
\end{equation}
by a three parameter correlated $\chi^2$-fit with $M=m_\sigma$. For
each fit, the errors were obtained by varying $\chi^2$ by 1.

To measure the energy of the lowest laying state in the $I=1$, $J=1$ 
channel, an operator carrying
these quantum numbers needs to be constructed. The simplest such
operator is:
\begin{equation}
O_{c,m}(t)={1\over{L^3}} \sum_{x \in \Lambda_t}
\pi_x^a \pi_{x+\hat m}^b \epsilon_{abc}
\end{equation}
where summation over repeated indices is assumed, 
$a, b, c $ are the isospin indices, 
$\epsilon_{abc}$ is the totally antisymmetric tensor,
$\hat m$ is the $m$'th
Euclidean unit vector of the time slice $\Lambda_t$, and $m \in
[1,2,3]$ is the z-component spin index.  Unfortunately, the time slice
connected correlation function of this operator gives a very weak
signal.  To get a better signal an operator that extends over several
lattice spacings needs to be constructed using a trial wave function
for the two--pion state. The ``bag'' and ``bound state'' meson wave
functions were considered \cite{DeGrand&Loft}. The latter gave a
better signal with an affordable cost in computer time and was
therefore used for the simulation.  Using as a trial wave function for
the two pions at relative position $\vec R$, the hydrogen wave
function $\Psi_{n=2,l=1,m}(\vec R)$, an $I=1$, $J=1$ operator with total
3-momentum zero was constructed:
\begin{equation}
O_{c,m}(t)=\sum_{x \in \Lambda_t}\left\{ \sum_{\vec R \in B}
|\vec R| \exp({-|\vec R|/2a_0}) \ Y_{1,m}(\theta,\phi) 
\pi_x^a \pi_{x+\vec R}^b \epsilon_{abc}
\right\}
\label{rho_op}
\end{equation}
where $B$ is a three-dimensional cubic box centered at the origin and
contained in $\Lambda_t$, $a_0$ is the ``Bohr radius'' in lattice
units, $\theta$ and $\phi$ are the azimuthal and polar angles of $\vec
R$, and $Y_{1,m}$ is the $l=1$ spherical harmonic (up to a
multiplicative normalization constant).  The parameter $a_0$ can be
given any value. A very large $a_0$ will cause the exponential to
decrease very slowly and then the sum over $\vec R$ will have to be
carried out over a box $B$ as large as $\Lambda_t$.  Since this can
increase the computer time significantly, a smaller $a_0$ has to be
used so that the size of the box that contains the important
contribution from the exponential can be made smaller. However, $a_0$
cannot be made very small because the signal becomes weaker as $a_0$
decreases. An optimal choice of $a_0$ and $B$ was found to be $a_0=2$,
and $B$ extending from $-3$ to $+3$ in each of the three directions.
The operator $O_{c,m}(t)$ couples to the $I=1$, $J=1$ states.  The energy
of the lowest laying state in this channel can be found by looking at
the time slice connected correlation function 
$C_{c,m}(t)=<O_{c,m}(0)O_{c,m}(t)^*>_c$ of this operator.

The simulation was done on an $L=8, L_t=16$ and $L=16, L_t=16$
lattice and for three values of the hopping
parameter $\kappa=0.305, 0.310, 0.330$. These values were chosen
so that a wide range of
$m_\sigma$ will be covered (they also coincide with some of
the values used in \cite{Has}). The expectation values of 
$C_{0,0}(t)$, $C_{1,1}(t)={1\over 9}\sum_{c,m} C_{c,m}(t)$,
and also of the pion zero 3--momentum time slice correlations 
were measured. The pions
will not concern us here except to mention that they are 
the massless Goldstone modes and found to behave as in \cite{Has}. 

The number of sweeps, the number of measurements of
$C_{0,0}(t)$ and $C_{1,1}(t)$, and the
the values of $m_\sigma$, for the various lattice sizes and $\kappa$'s,
are given in table 1. The results for $L=16$ are in good
agreement with the results of \cite{Has}.

\begin{table}
\begin{center}
\begin{tabular}{||c|c|r|r|l||}   \hline
$\kappa$ & L  & sweeps $\ $  & measurements &$\ \ \ m_\sigma $ \\ \hline\hline
0.305    & 16 &   4$\times 10^5$ & 1.6$\times 10^4$& 0.225(4)    \\ \hline 
0.310    & 16 & 1.6$\times 10^5$ & 1.6$\times 10^4$& 0.424(6)    \\ \hline 
0.330    & 16 & 0.8$\times 10^5$ & 1.6$\times 10^4$&  0.83(1)    \\ \hline
0.305    &  8 &17.5$\times 10^5$ &   7$\times 10^4$& 0.343(2)    \\ \hline 
0.310    &  8 &   7$\times 10^5$ &   7$\times 10^4$& 0.468(3)    \\ \hline 
0.330    &  8 &   1$\times 10^5$ &   2$\times 10^4$&  0.91(2)    \\ \hline
\end{tabular}
\end{center}
\caption{Number of measurements and $m_\sigma$.}
\medskip
\begin{center}
\begin{tabular}{||c|c|l|c|c||}  \hline
$\kappa$ & L  & $\ \ \ E$ & fit--range & $\chi^2/$d.o.f.\\ \hline\hline
0.305    & 16 & 0.94(1)   &   2--8     &     8.8        \\ \hline
0.305    & 16 & 0.76(3)   &   3--8     &     1.8        \\ \hline
0.310    & 16 & 0.94(1)   &   2--8     &     0.7        \\ \hline
0.310    & 16 & 0.91(3)   &   3--8     &     0.7        \\ \hline
0.330    & 16 & 0.94(1)   &   2--8     &     4.2        \\ \hline
0.330    & 16 & 0.87(2)   &   3--8     &     2.4        \\ \hline
0.305    &  8 & 1.549(5)  &   1--5     &     1.7        \\ \hline
0.310    &  8 & 1.531(6)  &   1--5     &     1.7        \\ \hline
0.330    &  8 & 1.501(9)  &   1--4     &     0.4        \\ \hline
\end{tabular}
\end{center}
\caption{Energy $E$ in the $I=1$, $J=1$ channel.}
\end{table}

$C_{1,1}(t)$ is plotted versus $t$ in figure 1 for the $L=16$ lattice
and the three values of $\kappa$. 
$C_{1,1}(t)$ was fitted with the expression in 
equation \ref{cosh} with $b=0$ for a few different ranges of $t$, and the
resulting energies $E=M$, together with the $\chi^2$ per degree of
freedom for each fit are given in
table 2 for the $L=16$ and $L=8$ lattices. The dotted line in figure 1
is the fit for $\kappa=0.310$, $L=16$, and $2 \le t \le 8$ and is presented
to give the reader a feeling of the quality of the fits.

\epsfxsize=4.8 in
\centerline{\epsffile{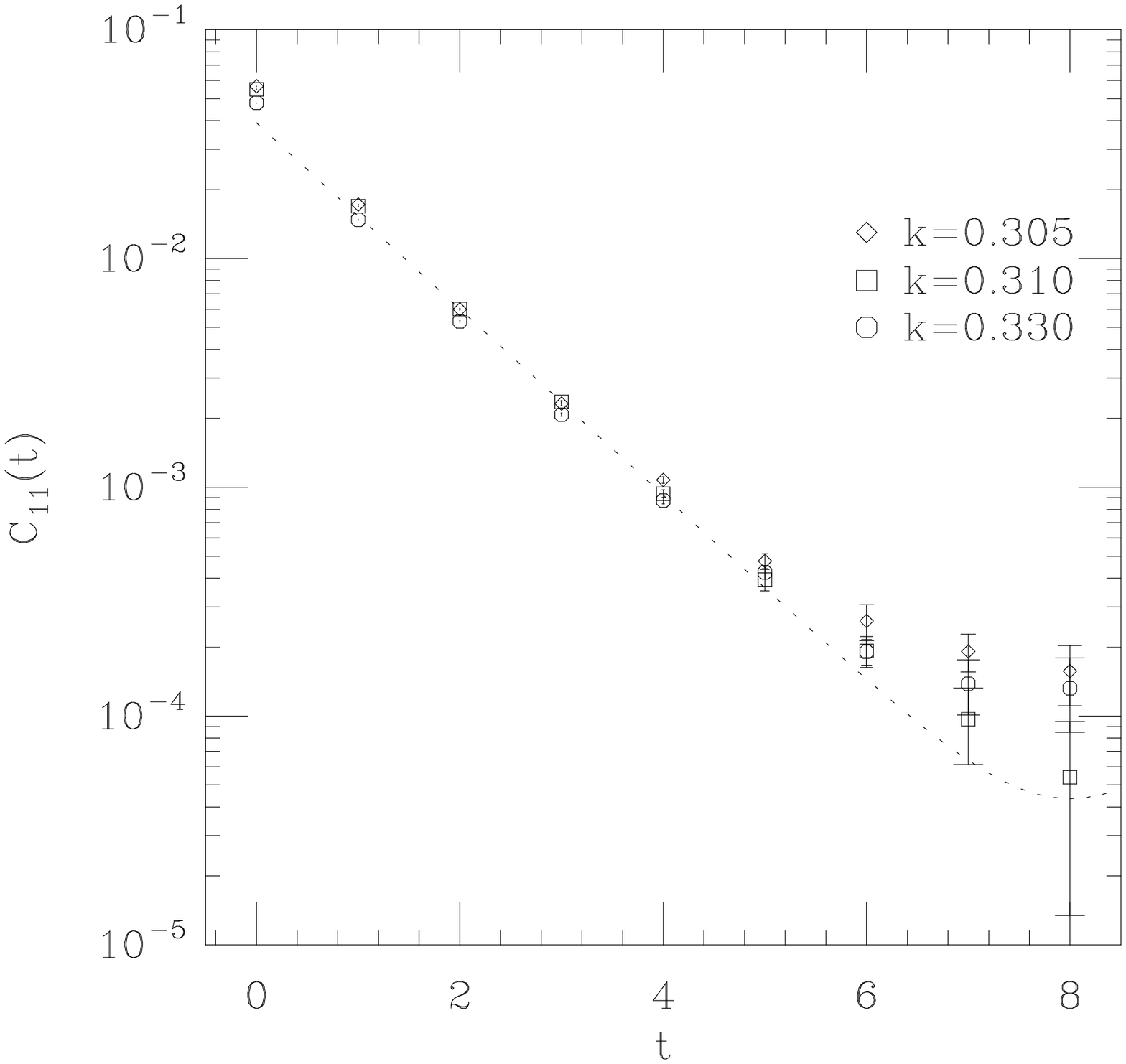}}
\centerline{{\bf Figure 1:}$\ < C_{1,1}(t)>$ for the $L=16$, $L_t=16$ lattice.}
\medskip

The effective energy $E_{\rm eff}(t)$, obtained by fitting 
$C_{1,1}(t)$ to the expression in equation \ref{cosh}, with $b=0$ for the two
time slices at t-1 and t,
is plotted versus t and for the three
values of $\kappa$ in figures 2 ($L=16$) and 3 ($L=8$).
The values of t
omitted from those plots had an $E_{\rm eff}(t)$ with very large error.

In a two-pion $I=1$, $J=1$ state with zero total 3--momentum,
the lowest 3--momentum a pion can have has one component
equal to $2\pi\over L$ and two equal to $0$.  The next one has two
components equal to $2\pi\over L$ and one equal to $0$. The energy
spectrum in the $I=1$, $J=1$ channel is therefore expected to contain
levels with energies close to the energies of these states, but
slightly different because of the interaction. For the $L=16$ lattice,
these levels have energies
$E_0\simeq 0.78$ and $E_1\simeq 1.09$
respectively, and are denoted by the dotted lines 

\epsfxsize=4.8 in
\centerline{\epsffile{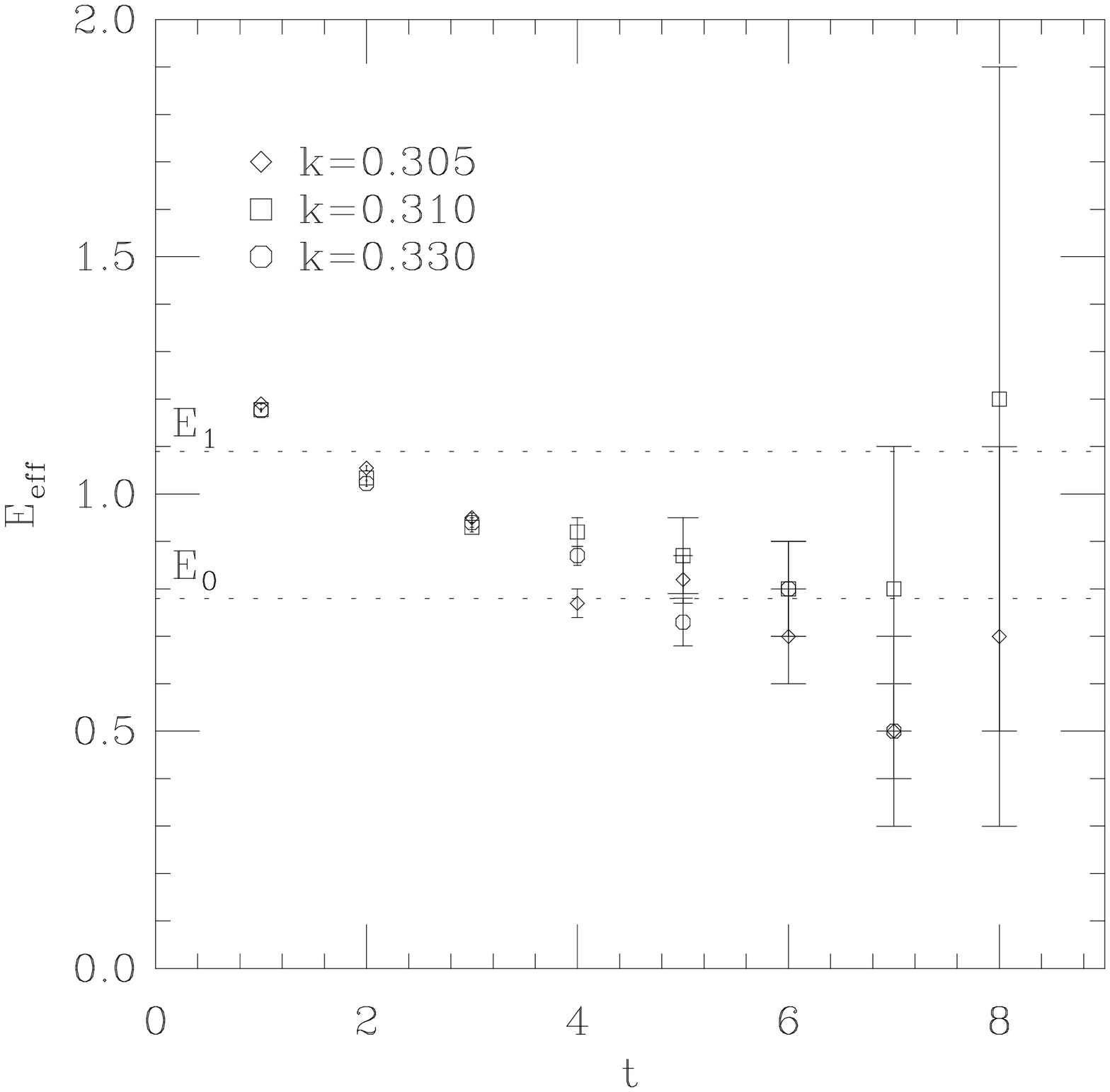}}
\centerline{{\bf Figure 2:}$\ E_{\rm eff}(t)$ for the $L=16$, $L_t=16$ lattice.}
\medskip

\noindent 
in figure 2.  From
this figure, it is clear that the observed energy levels are very
close to the levels of two free pions.  In fact, for smaller $t$, the
levels are close to $E_1$, and for larger $t$ they are close to $E_0$.
Because the free two--pion levels are not very well separated at
$L=16$, the observed levels are probably a mixture of the two lowest
ones. In that sense, the energies in table 2 for
the $L=16$ lattice are probably a mixture as well. The fact that the
observed levels
correspond to a two--pion state and not to a resonance is also greatly
supported by the fact that these levels change only slightly from
$\kappa=0.305$ to $\kappa=0.330$.  After all, in that range $m_\sigma$
varies from $0.225$ to $0.91$. Therefore, if a resonance is present 
it must have energy larger than $\approx 0.78$ and is either too heavy
(for example, heavier than $1.09$) to be observed with this method,

\epsfxsize=4.8 in
\centerline{\epsffile{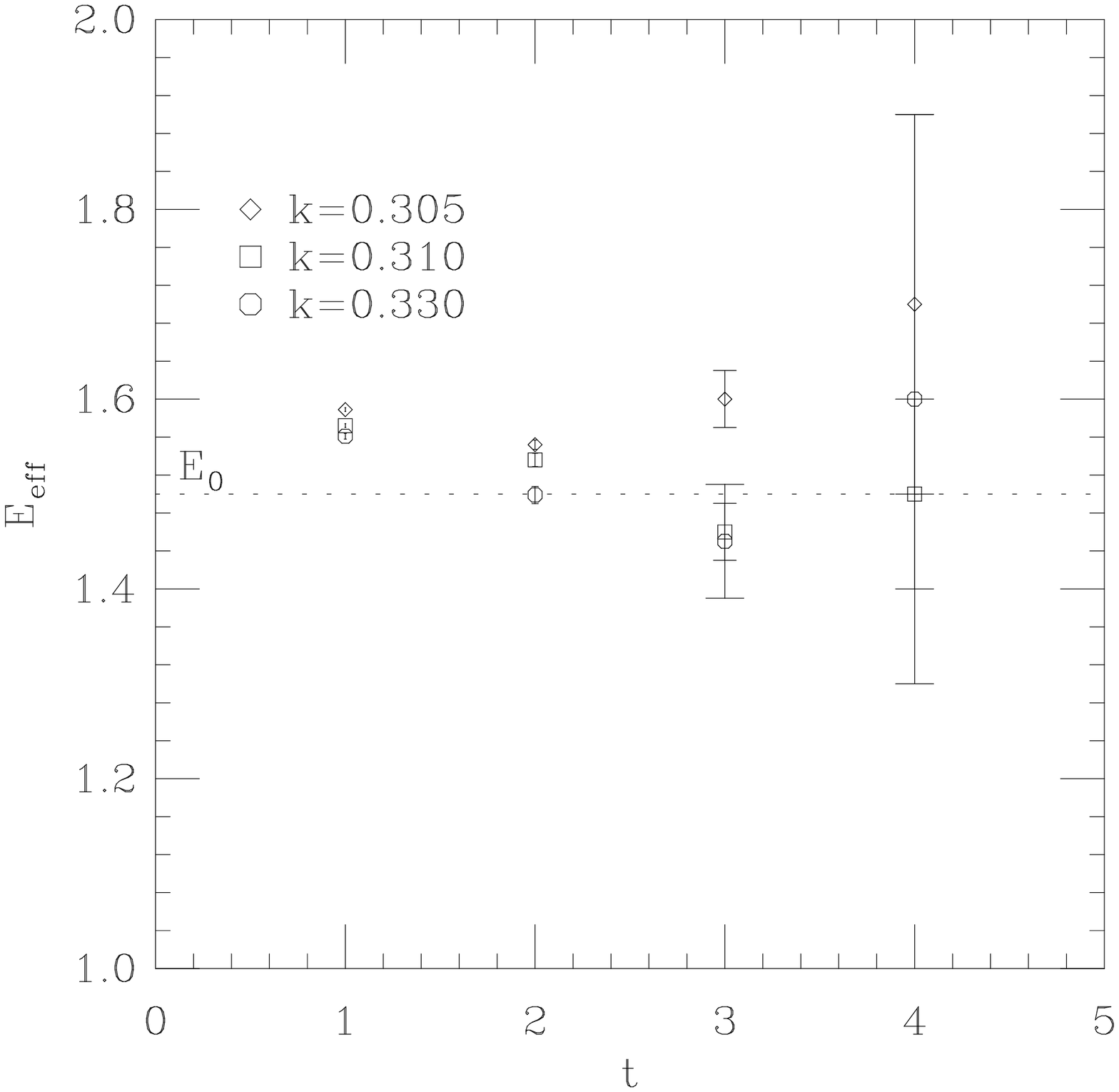}}
\centerline{{\bf Figure 3:}$\ E_{\rm eff}(t)$ for the $L=8$, $L_t=16$ lattice.}
\medskip

\noindent or is ``hiding'' between $0.78$ and $1.09$. That the latter
is not the case can be seen by looking at the energy levels obtained
for the $L=8$ lattice. From figure 3 it is clear that there are no
energy levels below $\approx 1.4$. In fact, since the two lowest free
two--pion states are well separated in this lattice, the lowest energy
level is clearly visible in figure 3.  It is true that the limited
statistics give $E_{\rm eff}(t)$ only up to $t=4$, but because the lowest
level is well separated from the next one, the correlated $\chi^2$ fit
gives a good estimate of the energy of this level. The numbers given
in table 2 are indeed very close to the lowest free two--pion energy
$E_0\simeq 1.50$ and do not seem to change much for the different
values of $\kappa$. From this analysis it is evident that if an $I=1$,
$J=1$ resonance exists for $\kappa \ge 0.305$, it is unphysical since
it must have energy in lattice units greater than $1$ (physical energy
above the cutoff).

For the theory of Gell-Mann and Levy of low energy pion processes the
$\kappa \ge 0.305$ region corresponds to where the $\sigma$ particle
mass is greater than approximately $180 MeV$ (or equivalently the
cutoff is less than approximately $1.3 GeV)$.\footnote{ The connection
with the physical scale $f_\pi$ (or $f_G$) is made through the
renormalized coupling constant $g_R=3m_\sigma^2/f_\pi^2$. The value of
$g_R$ for this value of $\kappa$ was taken from \cite{Luscher&Weisz}.}
Since $m_\sigma$ is expected to be much larger than $180 MeV$, it is
clear from the results that the existence of the  $\rho$ resonance in
nature cannot be accounted for by this low energy theory alone (in
contrast to \cite{Basdevant&Lee}).

For the scalar sector of the Minimal Standard Model the $\kappa \ge
0.305$ region corresponds to where the Higgs mass is greater than
approximately $500 GeV$ (or equivalently the cutoff is less than
approximately $3.5 TeV$). The Higgs mass is of course not known but
its upper bound is placed at around $650 GeV$ (hypercubic lattice
triviality bound). Thus the existence of an $I=1$, $J=1$ resonance can
be excluded with confidence for values of the Higgs mass above
$\approx 500 GeV$. For $\kappa< 0.305$ (Higgs mass $< 500 GeV$) the
strength of the interaction becomes weaker and hence it is safe to say
that if a resonance could not form for $\kappa \ge 0.305$, where the
interaction is stronger, it is unlikely that it will in this region
either. For this reason it was not deemed necessary to investigate the
$\kappa< 0.305$ region. A numerical simulation for $\kappa< 0.305$ not
only is not necessary, but it would also be very costly since larger
lattices will have to be used (the correlation length becomes larger
than approximately ${1\over 0.225}\simeq 4.5$).

It must be emphasized that these conclusions are valid only within the
realm of the scalar sector of the Minimal Standard Model.  It is of
course still  possible that the physics that enters at higher energies
may be able to produce such a resonance in very much the same way the
physical $\rho$ particle owes its existence to QCD. This resonance if
it exists due to some higher energy theory it would have energy
determined by that theory. In fact, it is possible that the energy of
this resonance is determining the cutoff energy of the Minimal
Standard Model. 

\medskip
\noindent{\bf Acknowledgements:} We would like to thank U.M. Heller for
his help and enlightening discussions. One of us (K.M.B.) would like
to thank A. Sanda for discussions concerning the subject of this
paper. This research was partially funded by the U.S.
Department of Energy through Contract No. DE-FC05-85ER250000.


\begin{thebibliography}{9}
\bibitem{Gell--Mann&Georgi} M. Gell--Mann, M. Levy, {\bf Il Nuovo
Cimento Vol. XVI, N. 4} (1960) 705; H. Georgi, {\bf Weak
Interactions and Modern Particle Theory} (1984) Addison-Wesley Pub.
\bibitem{Basdevant&Lee} See for example: J.L. Basdevant, B.W. Lee 
{\bf Phys. Rev. D Vol. 2, No 8} (1970) 1680; LH. Chan, R.W. Haymaker {\bf
Phys. Rev. D Vol. 10, No. 12} (1974) 4170.
\bibitem{equivalence} M.S. Chanowitz, M.K. Gaillard
{\bf Nucl. Phys.  B261} (1985) 379.  
\bibitem{Fredenhagen&Marcu} K.
Fredenhagen, M. Marcu {\bf Phys. Lett.  B 193} (1987) 486.
\bibitem{Has} A. Hasenfratz, K. Jansen, J. Jersak, C.B. Lang, T.
Neuhaus, H. Yoneyama {\bf Nucl. Phys. B317} (1989) 81.
\bibitem{DeGrand&Loft} T.A. DeGrand, R.D. Loft {\bf Comp. Phys.
Comm.  65} (1991) 84.  
\bibitem{Luscher&Weisz} M. L\"uscher, P. Weisz
{\bf Nucl.  Phys. B318} (1989) 705.  
\end{thebibliography}
\end{document}